\documentclass[leqno,draft,11pt,a4paper]{article}
\usepackage{amssymb,amsmath,latexsym,theorem}
\usepackage[includeheadfoot,margin=1in]{geometry}

\usepackage{algorithm}
\usepackage[noend]{algpseudocode}

\def\sameenum{}
\def\shortthm{}
\ifx\mydefs\undefined\else \fi
\let\mydefs\relax

\ifx\slide\undefined
  \allowdisplaybreaks[2]
\fi

\ifx\loadcyr\undefined\else
\input cyracc.def
\makeatletter
 at 1\@ptsize pt
 at 1\@ptsize pt
 at 1\@ptsize pt
\makeatother

\fi

\def\gobble#1{}
\def\fixsup#1#2{{#1\let\dp\gobble\mathstrut}^#2_}

\def\bme{\hskip.75em\relax}

\def\delim#1#2#3#4{\ifx X#3X\left#1#4\right#2\else\csname#3l\endcsname#1#4\csname#3r\endcsname#2\fi}

\def\?{\mathbin?}

\newbox\circlebox
\setbox\circlebox\hbox{$\bigcirc$}
\def\circled#1{%
  \setbox0\hbox to\wd\circlebox{\hss$#1$\hss}\wd0=0pt
  \box0\copy\circlebox}

\let\tet\vartheta
\let\ep\varepsilon

\ifx\slide\undefined
  \def\greek#1{$\expandafter\greeknum\csname c@#1\endcsname$}
\else
  \def\greek#1{$\mathop{\boldsymbol{\expandafter\greeknum\csname c@#1\endcsname}}$}
\fi
\makeatletter
\def\greeknum#1{\ifcase#1\or\alpha\or\beta\or\gamma\or\delta\or\ep
      \or\digamma\or\zeta\or\eta\or\tet\or\iota\else\@ctrerr\fi}
\makeatother

\def\p#1{\langle#1\rangle}

\def\lh#1{\lvert#1\rvert}

\let\sset\subseteq

\let\onto\twoheadrightarrow

\newcommand\rpair[3][3em]{\mathrel{%
   \begin{matrix}%
     \strut\smash{\xrightonto{\hbox to#1{\hss$#2$\hss}}}\\[-1.7ex]%
     \strut\smash{\xleftembed[\hbox to#1{\hss$#3$\hss}]{}}%
   \end{matrix}}}
\makeatletter
\newcommand\xrightonto[2][]{\ext@arrow 0359\rightontofill{#1}{#2}}
\newcommand\xleftembed[2][]{\ext@arrow 3095\leftembedfill{#1}{#2}}
\def\leftembedfill{\arrowfill@\leftarrow\relbar\hookleftnoarrow}
\def\rightontofill{\arrowfill@\relbar\relbar\onto}
\def\hookleftnoarrow{\DOTSB\relbar\joinrel\rhook}
\makeatother

\def\fl#1{\lfloor#1\rfloor}

\mathchardef\#="2023 

\ifx\busspf\undefined\else
\usepackage{bussproofs}
\EnableBpAbbreviations

\fi

\ifx\symlasy\undefined

  \def\docentdot#1#2{%
    \setbox0\hbox{$#1\mathop{#2}$}\dimen0 \ht0
    \setbox0\hbox{$#1#2$}\advance\dimen0 -\ht0
    \setbox2\hbox to\wd0{\hss$#1\mathop{\cdot}$\hss}\wd2=0pt
    \lower\dimen0\box2\box0 }
\else
  
  \def\docentdot#1#2{%
     \setbox0\hbox{$#1#2$}%
     \raise0.206\ht0\hbox to\wd0{\hss$#1\cdot$\hss}%
     \kern-\wd0 \box0 }
\fi

\def\Up{{\setbox0\hbox{$\uparrow$}%
         \lower\dp0\hbox to\wd0{\hss\vrule width4pt height.4pt\hss}%
         \kern-\wd0\box0}}
\def\UP{{\setbox0\hbox{$\uparrow$}%
         \lower\dp0\hbox to\wd0{\hss\vrule width4pt height.4pt\hss}%
         \kern-\wd0\copy0\kern-\wd0\raise.35ex\box0}}
\def\Down{{\setbox0\hbox{$\downarrow$}%
         \raise\ht0\hbox to\wd0{\hss\vrule width4pt depth.4pt\hss}%
         \kern-\wd0\box0}}

\newif\ifnadm

\def\doadm{\mathrel{%
   \setbox0 \hbox{$\mathop\vdash$}\dimen0 \ht0
   \setbox0 \hbox{$\vdash$}\advance\dimen0 -\ht0
   \vrule width.8\fontdimen8 \textfont3 height\ht0 depth\dp0
   \mkern-1mu
   \lower\dimen0 \hbox{$\vcenter{%
      \ifnadm
        \setbox0 \hbox{$\scriptstyle\sim\mathstrut$}%
        \hbox{\hbox to\wd0{\hss$\scriptstyle/$\hss}\kern-\wd0 \box0 }%
      \else
        \hbox{$\scriptstyle\sim\mathstrut$}%
      \fi}$}}}

\def\nrstyle#1#2#3{%
  \setbox0\hbox{$#2\bigcirc$}%
  \vcenter{\hbox to\wd0{\hss$#3#1$\hss}}%
  \kern-\wd0\box0 }

\ifx\slide\undefined
  
\fi

\def\task#1{{\normalfont\textsc{#1}}}

\let\thry\mathsf

\def\st{\expandafter\hat}

\def\apc#1{\expandafter\doapc\expandafter{\expandafter\fam\the\fam\relax#1}}
\def\doapc#1{\thry{APC_{#1}}}

\def\pfsys#1#2{\ifx\relax#2\relax\else#2\text-\fi\mathrm{#1}}

\def\doindep#1#2{\mathrel{\mathop{\vcenter{%
        \hbox{\oalign{\noalign{\kern-.3ex}\hfil$#1$\hfil\cr
              \noalign{\kern-.7ex}%
              $\smile$\cr\noalign{\kern-.3ex}}}}}\displaylimits_{#2}}\nobreak}
\def\nvert{{\setbox0\hbox{$\smallsetminus$}\hbox to\wd0{\hss$\vert$\hss}\kern-\wd0\reflectbox{\box0}}}

\mathcode`\*="0203

\mathchardef\mhyphen="2D
\def\cput(#1)#2{\put(#1){\hbox to0pt{\hss#2\hss}}}

\makeatletter
\def\iddots{\mathinner{\mkern1mu\raise\p@\hbox{.}\mkern2mu
        \raise4\p@\hbox{.}\mkern2mu\raise7\p@\vbox{\kern7\p@\hbox{.}}\mkern1mu}}
\makeatother

\ifx\slide\undefined

\def\noproof{\leavevmode\unskip\bme\vadjust{}\nobreak\hfill$\qed$\par}
\let\qed\Box
\newenvironment{Pf}[1][]
  {\par\noindent\textit{Proof\optpar{#1}:}\bme\ignorespaces}
  {\noproof\pagebreak[2]\vskip\medskipamount\ignorespacesafterend}
\def\optpar#1{\ifx\relax#1\relax\else\ #1\fi}
\def\qedhere{\relax\ifmmode\eqno\qed\expandafter\aftergroup
                   \else\noproof\fi\noqed}
\def\noqed{\let\noproof\relax}

\theoremstyle{plain}
\ifx\shortthm\undefined
\newtheorem{Thm}{Theorem}[section]
\else
\newtheorem{Thm}{Theorem}
\fi

\ifx\shortthm\undefined
\ifx\hiercl\undefined
\def\theCl{\arabic{Cl}}
\fi\fi

\theorembodyfont\upshape

\newenvironment{Pf*}{\let\qed\qedCl\Pf}\endPf

\fi

\ifx\slide\undefined
  \usepackage[reftex]{theoremref}
\fi
\makeatletter
\def\thmref@flush{%
   \ifx\thmref@last\empty\else
      \ifthmref@comma, \thmref@finaltrue\fi \thmref@commatrue
      \thmref@last \ifx\thmref@stack\empty\else s\fi \thmref@num 0
      \let\do\thmref@one \thmref@stack
      \ifcase\thmref@num\or\space and\else\thmref@finaltrue, and\fi
      ~\ref{\thmref@head}\let\thmref@stack\empty\fi}
\def\thmref@one#1{\ifnum\thmref@num>0,\fi
   \space\ref{#1}\advance\thmref@num 1\relax}
\makeatother

\begingroup \lccode`\~=`\/ \lowercase{ \endgroup
  \providecommand\url{\begingroup \catcode`\~=12 \catcode`\_=12 \catcode`\/=13 \let~\beginslash \finishurl}
  \def\finishurl#1{\texttt{#1}\endgroup}
  \def\beginslash{/\futurelet\nexttoken\finishslash}
  \def\finishslash{\ifx\nexttoken~\else\penalty\relpenalty\fi}
}

\newif\iflinenumbers
\linenumberstrue

{\catcode`\^^I=13 \catcode`\^^M=13
\gdef\doalgo#1#2\end#{\hbox to\hsize{\hss \let^^I\qquad%
  \def\\^^M{\nobreak\hfil\break\vadjust{}\qquad}%
  \def\>{{\everypar{}\indent}}%
  \def\<{{\everypar{}\noindent}}%
  \fboxsep1em \linenum0 \parindent1.2em%
  \fbox{\hsize#1\vbox{%
  \everypar{\iflinenumbers\advance\linenum1 \llap{$\scriptstyle\the\linenum$\hskip.6em}\fi}%
  #2}}\hss}\end}}
\newcount\linenum

\def\key{\relax\ifmmode\expandafter\mathbf\else\expandafter\textbf\fi}
\def\proc{\relax\ifmmode\expandafter\mathrm\else\expandafter\textrm\fi}

\def\allowhyphens{\nobreak\hskip0pt\relax}

\DeclareRobustCommand*\magiclparen{\ifmmode(\else\textup(\allowhyphens\fi}
\DeclareRobustCommand*\magicrparen{\ifmmode)\else\textup)\fi}
\let\lparen=(  \let\rparen=)
\def\magicparon{\catcode`\(\active\catcode`\)\active}
\def\magicparoff{\catcode`\(12 \catcode`\)12 }
\AtBeginDocument{\magicparon}
\magicparon
\let (=\magiclparen  \let )=\magicrparen

\ifx\sameenum\undefined
  \def\theenumi{\roman{enumi}}
  \ifx\enumup\undefined
    
  \else
    
  \fi

\else
  \makeatletter
  \ifx\enumup\undefined
     \def\theenumi{\ifnum\@enumdepth=\@ne(\roman{enumi})\fi}
  \else
     \def\theenumi{\ifnum\@enumdepth=\@ne\textup{(\roman{enumi})}\fi}
  \fi
  \makeatother

\fi

\magicparoff

\mathchardef\comma=\mathcode`\,
{\catcode`\,=\active \gdef,{\comma\penalty\relpenalty}}

\ifx\slide\undefined

\providecommand\dedic{%
  \message{^^JWARNING: embed explicit dedication in the paper!^^J}%
  \thanks{Supported by the Czech Academy of Sciences (RVO 67985840)
      and GA \v CR project 23-04825S.}}

\author{Emil Je\v r\'abek\dedic\\[\medskipamount]
Institute of Mathematics, Czech Academy of Sciences\\
\small \v Zitn\'a 25,
115\:67 Praha 1,
Czech Republic,
email: \texttt{jerabek@math.cas.cz}
}

\else\ifx

\author{Emil Je\v r\'abek}
\email{jerabek@math.cas.cz\\[-1em]https://math.cas.cz/\string~jerabek/}
\institution{Institute of Mathematics, Czech Academy of Sciences}

\else 

\author{Emil Je\v r\'abek}
\institute{Institute of Mathematics\\Czech Academy of Sciences\\\texttt{jerabek@math.cas.cz}\\\texttt{https://math.cas.cz/\string~jerabek/}}

\fi\fi

\def\ssum{\task{SeqSum}}
\DeclareMathOperator\divmod{divmod}
\def\head#1{#1.\mathrm{head}}
\def\mvl#1{#1.\mathrm{left}}

\def\dedic{\thanks{Supported by the Czech Academy of Sciences (RVO 67985840) and GA \v CR project 23-04825S.}}

\title{A note on the complexity of addition}

\begin{document}
\maketitle

\begin{abstract}
We show that the sum of a sequence of integers can be computed in linear time on a Turing machine. In particular, the
most obvious algorithm for this problem, which appears to require quadratic time due to carry propagation, actually
runs in linear time by amortized analysis.

\smallskip
\noindent\textbf{Keywords:} computational complexity, integer addition, amortized analysis
\end{abstract}

\section{Introduction}\label{sec:introduction}

Elementary arithmetic operations on integers are some of the most basic algorithmic tasks, and their computational
complexity is of fundamental importance. Even for such simple problems, accurate determination of their asymptotic time
complexity (always measured on \emph{multi-tape Turing machines (TM)} in this note) may be surprisingly difficult:
while the complexity of $+$, $-$, and $<$ is clearly $\Theta(n)$ (the school-book algorithms run in time $O(n)$, which
is optimal as any algorithm must at least read the whole input), the complexity of $\times$ and~$/$ is still not
settled after decades of research---the recent $O(n\log n)$ upper bound due to Harvey and van der Hoeven~\cite{hh:mult}
is regarded as optimal by many, but no nontrivial lower bounds are known\footnote{%
Assuming the network coding conjecture, Afshani et~al.\ \cite{afkl:mult} proved an $\Omega(n\log n)$ bound on the
number of wires in multiplication circuits, but besides being conditional, this does not imply any super-linear lower
bounds on TM time complexity.}.

The problem we are interested in in this note is nowhere near as difficult to analyze, however it is just as basic, and
yet it appears to have been missed by standard literature (cf.~\cite{ej:sum-q}). We consider the generalization of $+$
to summation of \emph{many} integers: that is, the input is a sequence $\p{X_i:i<k}$ of nonnegative integers written in
binary and separated by $+$ signs (say), and the output is $\sum_{i<k}X_i$. (All our indices start from~$0$.) We denote
this problem $\ssum$. The size of the input is $\max\{k-1,0\}+\sum_in_i$, where $n_i=\lh{X_i}$ denotes the length
of~$X_i$ in binary; in order to simplify bounds, we will use the slightly different parameter $n=k+\sum_in_i$. Can we
still compute the result in time $O(n)$?

\section{The accumulator algorithm}\label{sec:accum-algor}
\begin{algorithm}[t]
\caption{The accumulator algorithm}
\label{alg:accu}
\begin{algorithmic}[1]
\Procedure{SeqSum}{$X$,$Y$}
  \While{$\head X\ne\hbox{\textvisiblespace}$}
    \State $c\gets0$ \Comment{carry bit}
    \While{$\head X\in\{\texttt0,\texttt1\}$} \Comment{addition loop}
      \State $\p{c,\head Y}\gets\divmod(\head X+\head Y+c,2)$ \label{ln:5}
      \State $\mvl X,\mvl Y$ \label{ln:6}
    \EndWhile
    \While{$c\ne0$} \Comment{carry propagation}
      \State $\p{c,\head Y}\gets\divmod(\head Y+c,2)$
      \State $\mvl Y$      
    \EndWhile
    \State $\mvl X$ \Comment{skip separator}
    \State rewind $Y$ back to the right end \label{ln:11}
  \EndWhile
\EndProcedure
\end{algorithmic}
\end{algorithm}
Alg.~\ref{alg:accu} describes a baseline TM algorithm for $\ssum$, which uses one tape as an
``accumulator'' to which we add the input numbers $X_i$ one by one. In the pseudocode, $X$ denotes the input tape, and
$Y$ the accumulator tape (which may be the output tape if we do not insist on its being write-only, or it may be a work
tape that is duplicated to the output). We assume the tapes are infinite to the left, and the machine starts with heads
at the right-most positions; the input sequence is written from right to left, with each integer starting with the
least-significant bit on the right (as humans are used to). The symbol under the head of a tape~$T$ is accessed with
$\head T$, and the $\mvl T$ instruction moves the head one position to the left. The carry bit $c\in\{0,1\}$ is
included in the TM state. The $\divmod(a,b)$ operator computes the pair $\p{\fl{a/b},a\bmod b}$; note that lines
\ref{ln:5} and~\ref{ln:6} together really denote a single step of the machine that updates $\head Y$ and~$c$ (i.e., the
state) based on the current contents of $\head X$, $\head Y$, and~$c$, and moves both tape heads. (See \cite{ej:sum-tm}
for a working implementation of the Turing machine, in an environment with somewhat different tape conventions.)

In order to analyze the time complexity of Alg.~\ref{alg:accu}, consider an input $\p{X_i:i<k}$, write $n_i=\lh{X_i}$
and $n=k+\sum_{i<k}n_i$ as in Sec.~\ref{sec:introduction}, and put $Y_i=\sum_{j<i}X_j$ and $m_i=\lh{Y_i}$. The $i$th
(counting from~$0$) iteration of the outer loop performs the addition $X_i+Y_i\to Y_{i+1}$, and takes time
$O(\max\{n_i,m_i\})\sset O(n)$ as $m_i\le n$; the overall running time is thus $O(nk)\sset O(n^2)$. While these
estimates are somewhat wasteful, there is no obvious way how to improve them. In particular, the addition of $X_i$
to~$Y_i$ may cost time $\approx m_i$ even if $n_i$ is much smaller than~$m_i$ due to carry propagation, and there are
instances where $m_i=\Omega(n)$ for all $i\ne0$, and $k=\Omega(n)$, whence $\sum_{i<k}m_i=\Omega(n^2)$: e.g., take
$\lh{X_0}=n/2$ and $X_i=1$ for $1\le i<k=n/4$. Thus, Alg.~\ref{alg:accu} appears to require quadratic time at first
sight.

Nonetheless, we will show that Alg.~\ref{alg:accu} only needs time $O(n)$ by a more refined argument.

\section{Amortized analysis}\label{sec:amortized-analysis}
We will estimate the running time of Alg.~\ref{alg:accu} more accurately using basic methods of amortized
complexity~\cite{tarj:amort}. The point is that even though a single addition $X_i+Y_i\to Y_{i+1}$ may cost time up to
$m_i$ due to long carry propagation, the latter cannot happen very often; but we need to quantify this idea properly to
see that it indeed works out to an $O(n)$ overall cost.

A well-known simple special case is the \emph{binary counter} (see e.g.~\cite[pp.\ 451\,ff.]{clrs}): the counter holds
an integer in binary, starting with~$0$, and we perform $n$ increment operations; this is virtually identical to
Alg.~\ref{alg:accu} applied to an input of the form $1+1+\dots+1$. Each increment may access up to $\log n$ bits, hence
a na\"\i ve bound on the running time is $O(n\log n)$. But while all $n$ increments access the $0$th bit of the
counter, only $n/2$ access the $1$st bit, $n/4$ the $2$nd bit, etc., thus the overall running time is only $O(n)$.

For a general instance of Alg.~\ref{alg:accu}, it seems difficult to directly count the number of carries in a similar
way. However, we can prove a linear upper bound using the ``accounting method''.
\begin{Thm}\label{thm:main}
Algorithm~\ref{alg:accu} solves $\ssum$ in time $4n+1$.
\end{Thm}
\begin{Pf}
For each $i<k$, let $t_i$ denote the furthest bit position (starting from~$0$) of~$Y$ accessed by Alg.~\ref{alg:accu}
during the addition of $X_i$ to the accumulator. This addition takes $2(t_i+1)$ steps (including the rewind on
line~\ref{ln:11}), hence the total running time\footnote{%
The final ${}+1$ is for a transition to a halting state; depending on the exact details of the formalization of TM, it
may be unnecessary. The reader may check that it can be shaved off anyway as long as $n\ge1$.} %
of the algorithm is $t=2\sum_i(t_i+1)+1$.

If there is no carry propagation, then $t_i=n_i$, and the addition loop modifies up to $n_i$ bits of~$Y$ in positions
$0,\dots,n_i-1$.

If there \emph{is} carry propagation, then $t_i\ge n_i$, the main addition loop may modify some bits in positions
$0,\dots,n_i-1$, and the carry propagation loop modifies positions $n_i,\dots,t_i$: the bit in position~$t_i$ changes
from $0$ to~$1$, and the $t_i-n_i$ bits in positions $n_i,\dots,t_i-1$ change from $1$ to~$0$. Since the accumulator
starts with~$0$, each of the latter $t_i-n_i$ changes can be uniquely associated with the latest change of $0$ to~$1$ in
the same position during the addition of $X_j$ to the accumulator for some $j<i$. We have already seen that the number
of such changes for a given $j$ is at most $n_j+1$, thus
\[\sum_{i<k}(t_i-n_i)\le\sum_{j<k}(n_j+1)=n.\]
It follows that
\[t=2\Bigl(\sum_{i<k}(n_i+1)+\sum_{i<k}(t_i-n_i)\Bigr)+1\le4n+1.\]

(In other words, we charge $2$~coins for each change $0\to1$, one of which is spent right away, and the other is saved
on that position as credit; changes $1\to0$ are then paid from credit.)

We can alternatively frame the argument using the ``potential method'': let $\Phi_i$ be the number of $1$s in $Y_i$.
The discussion above shows that $\Phi_{i+1}-\Phi_i\le n_i-(t_i-n_i)+1$, i.e.,
\[t_i+1\le2(n_i+1)+\Phi_i-\Phi_{i+1},\]
therefore
\[t\le4\sum_{i<k}(n_i+1)+2\sum_{i<k}(\Phi_i-\Phi_{i+1})+1=4n-2\Phi_k+1\le4n+1\]
as $\Phi_0=0$.
\end{Pf}

We mention that even though the problem is stated for binary integers for convenience, the result actually holds for
any base $b\ge2$. (There is also a trivial linear-time algorithm for unary integers.) Moreover, it generalizes to sums
of possibly negative integers, represented by a sign bit and absolute value: it suffices to sum the positive and
negative entries separately, and subtract the results.

Let us briefly discuss the role of the model of computation. The main alternative to TM in algorithmic complexity are
\emph{random-access machines (RAM)}. There is a lot of variation in RAM models both in terms of available features and
in terms of the definition of time complexity. Typically, simple arithmetic operations on registers such as addition
and subtraction are included as primitive operations. It follows that in unit-cost RAM models with unlimited register
size, the $\ssum$ problem is trivially solvable in linear time. If we limit register size to $O(\log n)$ bits (word
RAM) or in logarithmic-cost models, the problem becomes nontrivial again as the obvious algorithm needs quadratic time
(or at least appears so, as in the case of TM). We can still solve $\ssum$ in time $O(n)$ by Theorem~\ref{thm:main}
using an array of constant-size registers to simulate TM tapes, as long as we can access the individual entries of the
array with unit cost. But if indexing register number~$r$ costs $\log r$, this algorithm requires time
$\Theta(n\log n)$; in such a model, $\Theta(n\log n)$ time is necessary even just to read the input if it is also
represented as an array of bits.

\section{Conclusion}

We analyzed the performance of a straightforward algorithm for summing a sequence of integers on a TM, and found that it
takes time $O(n)$, despite first appearances suggesting it is only $O(n^2)$. The argument is quite simple, and it is an
extension of a standard example in amortized complexity (the binary counter). Nevertheless, to the best of our
knowledge it has not been observed before in commonly available literature; given the fundamental nature of the
problem, we believe it is worthwhile to point out the result explicitly.

\bibliographystyle{mybib}
\bibliography{mybib}

\providecommand\gobble[1]{} \ifx\url\undefined {\catcode`\/=13
  \gdef/{\string/\futurelet\nexttoken\finishslash}
  \gdef\finishslash{\ifx\nexttoken/\else\penalty\relpenalty\fi}}
  \def\url{\begingroup\catcode`\~=12 \catcode`\_=12 \catcode`\/=13 \finishurl}
  \def\finishurl#1{\texttt{#1}\endgroup} \fi \providecommand\href[2]{\url{#2}}
  \providecommand\dotminus{\mathbin{\scriptstyle\dot{\smash{\textstyle-}}}}
  \providecommand\hyph{\nobreak\hskip0pt-\hskip0pt\relax}
\providecommand\bysame{\leavevmode\hbox to5em{\hrulefill}\thinspace}
\providecommand\bibliographyhook{}
\begin{thebibliography}{1}
\bibliographyhook

\bibitem{afkl:mult}
Peyman Afshani, Casper~Benjamin Freksen, Lior Kamma, and Kasper~Green Larsen,
  \emph{Lower bounds for multiplication via network coding}, in: 46th
  {I}nternational {C}olloquium on {A}utomata, {L}anguages, and {P}rogramming
  ({ICALP} 2019) (C.~Baier, I.~Chatzigiannakis, P.~Flocchini, and S.~Leonardi,
  eds.), Leibniz International Proceedings in Informatics (LIPIcs) vol. 132,
  2019, pp.~10:1--12.

\bibitem{clrs}
Thomas~H. Cormen, Charles~E. Leiserson, Ronald~L. Rivest, and Clifford Stein,
  \emph{Introduction to algorithms}, fourth ed., MIT Press, Cambridge,
  Massachusetts, 2022, 1312~pp.

\bibitem{hh:mult}
David Harvey and Joris van~der Hoeven, \emph{Integer multiplication in time
  {$O(n\log n)$}}, Annals of Mathematics 193 (2021), no.~2, pp.~563--617.

\bibitem{ej:sum-q}
Emil Je{\v r}{\'a}bek, \emph{Can we do integer addition in linear time?},
  Theoretical Computer Science Stack Exchange, 2023,
  \url{https://cstheory.stackexchange.com/q/52391}.

\bibitem{ej:sum-tm}
\bysame, \emph{Sequence sum}, Martin Ugarte's Turing machine simulator, 2023,
  \url{https://turingmachinesimulator.com/shared/thnutsdfym}.

\bibitem{tarj:amort}
Robert~E. Tarjan, \emph{Amortized computational complexity}, SIAM Journal on
  Algebraic and Discrete Methods 6 (1985), no.~2, pp.~306--318.

\end{thebibliography}
\end{document}